\documentclass[useAMS,usenatbib,usegraphicx]{aa}  
\usepackage{natbib}
\usepackage{amsmath}

\usepackage{graphicx}
\usepackage{txfonts}
\begin{document}

   \title{Some remarks on the old problem of recombination}

   \author{ F. Jim\' enez Forteza
          \inst{1,3}
          \and
          J. Betancort Rijo  \inst{2,3}
          }

   \institute{  Departament de F\'isica, Universitat de les Illes Balears, Crta. Valldemossa km 7.5, E-07122 Palma, Spain\\
             \email{f.jimenez@uib.es}
            \and
             Instituto de Astrof\'isica de Canarias, E-38200 La Laguna, Tenerife, Spain\\
             	\email{jbetanco@iac.es}
         \and
         Departamento de Astrof\'isica, Universidad de La Laguna, E-38205 La Laguna, Tenerife, Spain
             }

   \date{Received *, **; accepted *, **}

% \abstract{}{}{}{}{} 
% 5 {} token are mandatory
 
  \abstract
   {In the seminal works of \citet{Zeldovich1968} and \citet{Peebles1968}, a procedure  was outlined to obtain the equation of evolution of the hydrogen fraction without an explicit use of the radiative transfer equation. This procedure is based, explicitly or implicitly, on the concept of escape probability and, using the Sobolev approximation for this problem \citep{Sobolev1960}, has extensively been used since then in developing refined approximations.}
   {To derive in a simple, rigorous and general manner the above mentioned procedure and
to obtain exact analytical expressions for the spectral density of radiation generated at one photon and two photon recombination transitions. These expressions are used to estimate
the implications of several interesting effects.}
   {Some slight re-elaborations of basic principles of transport theory.}
   {We have obtained the expressions searched for and used them in several explicit computations. We have found that the relative change in the electronic fraction due to the absorption by the two photon line, $1s\rightarrow 2s$, of photons that have escaped from the line $2p\rightarrow 1s$ is $0.67\%$, that is a result $12\%$ higher than the previous ones obtained using some approximations \citep{Kholupenko2006,Hirata2008}. The photons generated by the transition $2s\rightarrow 1s$ and later absorbed by the same transition (in combination with a photon more energetic than its original partner) imply a $0.05\%$ maximum variation of the electronic fraction. This problem has not been treated previously analytically, although a numerical estimate have been recently carried out in \citet{Chluba2011}. }
   {}

   \keywords{cosmic background radiation 
               }

   \maketitle
%
%________________________________________________________________

\section{Introduction.}
The initial aim of this work was to give an account of the basic recombination picture through the escape of photons generated by hydrogen transitions in an as rigorous and simplified manner as possible. To this end, we have computed the hydrogen formation rate (in the ground state) due to the escape of Lyman $\alpha$ photon by mean of the escape probability. This has been done by a number of people \citep{Zeldovich1968, Peebles1968, Seager2000} in slightly different presentations using the Sobolev escape probability. To deal with the hydrogen destruction through absorption of CMB photons, we advanced an expression for the hydrogen net generation rate which is simply equal to the generation rate through escaping Lyman $\alpha$ photons times one minus the ratio of the actual hydrogen number density and that given by equilibrium. The expression satisfies the necessary requirement that in equilibrium the net rate is zero but it's not easy to prove that this expression is exact at intermediate cases where equilibrium does not hold  but where the correction factor still differ substantially from one. 
\paragraph*{}
It could seem that the expression is obviously true: if all Lyman $\alpha$ photon escape, the net hydrogen generation rate would be equal to the difference between the number of transitions $2p\rightarrow1s$ and $1s\rightarrow2p$ per unit time and unit volume; if the escape probability is smaller than one, the net rate should be modulated by this probability. However, on closer scrutiny, this turns out to be not obvious at all. In particular, it's not transparent the meaning of the escape probability multiplying the hydrogen destruction rate. Fortunately we have found a way to frame the problem that allows a rigorous and simple manner to prove that expression. In fact, in \citet{Peebles1968} a derivation for this is provided, but it's not as simple and transparent as the one given here. On top of that, we have found that our approach to the problem makes it possible an exact treatment of the problem by mean of a single differential equation, even when lines with frequencies higher than Lyman $\alpha$ are considered. Using our procedure to make an exact computation is beyond our interest and the scope of this work is to illustrate the relevance of the correcting terms with respect to similar but not exact treatments \citep{Seager2000}: we consider in detail examples of two processes. One of the processes is the absorption in the Lyman $\alpha$ line of photons generated by transitions leading to more energetic photons which are then redshifted into the Lyman $\alpha$. The other processes concern the two photon transition $(2s\rightarrow 1s)$, that we have noted that can not properly be treated as a single line, as is usually done in the standard scenario. When properly treated, stimulated emission plays a non-negligible role. We have found that this point has also been treated by \citet{Chluba2006}, \citep{Kholupenko2006} and \citet{Hirata2008} . The other two photon processes that we consider are: the absorption by the line $2s\rightarrow1s$ of a photon generated by the transition $2p\rightarrow1s$ \citep{Kholupenko2006,Hirata2008} with frequency $\nu_\alpha$ and redshifted to $\nu$ when combined with a CMB photon with frequency $\nu'$ such that $\nu+\nu'=\nu_\alpha$ and transitions $1s\rightarrow 2s$ generated by one of the two photons emitted at an earlier $2s\rightarrow 1s$, and that collectively escaped from that line through redshifting, combined with a CMB photon so that the sum of the frequencies is equal to $\nu_\alpha$.

\section{Rigorous derivation of the basic equation.}
If all the hydrogen is formed through a Lyman $\alpha$ transition it's clear that the time derivative of the hydrogen comoving density in the ground state is given by the following effective equation:

\begin{equation}
\begin{aligned}
\frac{dn_{H_{1s}}}{dt}&=p_{2p,1s}n_{H_{2p}}A_{2p,1s}\left(1+ (e^{\frac{h \nu_\alpha}{k T}}-1)^{-1}\right) \\
& -(1-e^{\tau_{2p,1s}})H(z)\nu_{\alpha}{\frac{a^3(t) B(\nu_{\alpha})}{h \nu_\alpha } }
\end{aligned}
\label{nsrate}
\end{equation}

where $n_{H_{2p}}$ is also comoving, $\frac{a^3(t) B(\nu_\alpha)}{h \nu}$ is the comoving density of photons in a unit frequency interval, $a(t)$ is the scale factor, $p_{2p,1s}$ is the escape probability and where we have taken into account the spontaneous and stimulated emission. The rational for this expression is that an hydrogen is created for any photon escaping from the line towards smaller frequencies (the first term on the right in equation \eqref{nsrate}), while an hydrogen is destroyed by any photon redshifted on to the line. The number of transitions $1s\rightarrow2p$ per unit time and frequency interval is given by the number of photons being redshifted into the line per unit of time and volume:

\begin{equation}
\left(H(z)\nu_\alpha\right) \left(\frac{B(\nu_\alpha,T)}{h\nu_\alpha}\right)
\label{eq:bb}
\end{equation} 

multiplied by the probability for a photon to be absorbed while crossing the entire line (from the blue side to the red one), which in terms of the optical depth, $\tau$ can be written in the form:$$1-e^{-\tau_{2p,1s}}$$ 

Notice that expression \eqref{eq:bb} is simply the product of the photons ``speed'' in $\nu$-space (first parenthesis) by the $\nu$-space photon number density per unit volume (second parenthesis). We have written expression \eqref{nsrate} in terms of the general concepts denoted by $p$ and $\tau$, so that it can account for the fact that the emission and absorption profiles are not identical \citep{Chluba2009}. To compute $\tau$ one must use the real absorption profile (excluding coherent scattering) so that $$1-e^{-\tau}$$ is the ``kill probability''. On computing $p$ one must take into account that when an ``absorbed'' photon (including coherent scattering) is ``reemitted'' there is a superposition of real absorption and emission, with a frequency change distribution corresponding to the intrinsic width of the line plus the thermal widening and coherent scattering while for which the widening is essentially thermal. So, the ``absorption'' and ``emission'' profiles are not equal, although for real absorption and coherent scattering separately they are. Furthermore, two photon transitions from the $2p$ state to higher levels also contribute to widening the emission profile; this effect turns out to be the main effect in producing asymmetry between emission and absorption profiles \citep{Chluba2009_2}. In what follows, however, we shall not consider this fact, that is negligible as long as the thermal width it is much larger than the intrinsic one, and we use Sobolev  \citep{Seager2000} for $p_{(Sob)}$ and $\tau_{(Sob)}$ where the subscript $(Sob)$ refers to Sobolev values for probability and opacity. From this point, we use $p\equiv p_{(Sob) }$ and $\tau\equiv\tau_{(Sob)}$ for simplicity in the notation.
\paragraph*{}
Let us comment on the relationship between expression \eqref{nsrate} and the equation for the evolution of the number density of hydrogen (in the ground state) in an straightforward approach:

\begin{equation}
\label{eq:nsrate2}
\frac{dn_{H_{1s}}}{dt}=n_{H_{2p}}(A_{2p,1s}+B_{2p,1s}J(\nu_\alpha,T))- n_{H_{1s}} B_{1s,2p} J(\nu_\alpha,T)
\end{equation}

where $n_{H_{1s}}$, $n_{H_{2p}}$ are the comoving number densities and $J(\nu,T)$ is the spectral energy density of radiation, that it's made up of the initial black body radiation and that coming out from recombination lines, and whose evolution is given by the corresponding radiative transfer equation.
\paragraph*{}

In expression \eqref{nsrate} we make a different grouping of the terms than in \eqref{eq:nsrate2}. In \eqref{nsrate} we divide all photons in the line $\nu_\alpha$ in two categories: those generated by previous $2p\rightarrow1s$ transitions and those generated much earlier (when most of CMB photons were lastly generated before starting a merely passive evolution) and passively redshifted into $\nu_\alpha$. The transition involving the first category are accounted for by the first term in \eqref{nsrate}, while those in the second are accounted for by the second term. To compute this last term in \eqref{nsrate} it's essential the assumption that new photons getting trapped within the line (those already trapped are already included in the first term) are only those in the preexisting (before recombination) black body CMB (those due to higher frequency recombination lines will be considered later) being redshifted into the line. If the CMB were not evolving passively but there were processes generating new photons, an additional term should be included in \eqref{nsrate}. After some algebra, equation \eqref{nsrate} may be written in the form:

\begin{equation}
\label{eq:nsrate3}
\begin{aligned}
\frac{dn_{H_{1s}}}{dt}&=\frac{1-e^{-\tau_{2p,1s}}}{\tau_{2p,1s}}(n_{H_{2p}}A_{2p,1s}\left(1+(e^{\frac{h \nu_\alpha}{k T}}-1)^{-1}\right)\\
&-n_{H_{1s}}B_{1s,2p}B(\nu_\alpha))
\end{aligned}
\end{equation}

where $n_{H_{1s}}$, $n_{H_{2p}}$ are comoving densities. If neutral hydrogen were formed only through the two photon transition $2s\rightarrow1s$, an equation similar to \eqref{eq:nsrate3} would hold in as much as the transition can be described by a one photon model. The same holds for any other line if it were the only generating neutral hydrogen. When all lines contribute simultaneously, the equation of evolution for $n_{H_{1s}}$, in first approximation, has in the right hand side a sum of terms as that in \eqref{eq:nsrate3} (one for each line). 
\begin{equation}
\label{eq:n1s4}
\begin{split}
\frac{dn_{H_{1s}}}{dt}&=\sum_i \bigg(p_{i,1s}n_{H_{i}}A_{i,1s}\left(1+ e^{\frac{h \nu_\alpha}{k  T}}-1)^{-1}\right)+\\
& -p_{i,1s}n_{H_{1s}}B_{1s,i}B(\nu_i)\bigg)
\end{split}
\end{equation}

where for formal simplicity in the forthcoming equations we use the following net rate definition: $$C_i \equiv p_{i,1s}n_{H_{i}}\left(A_{i,1s}\left(1+(e^{\frac{h \nu_\alpha}{k T}}-1)^{-1}\right)-n_{H_{1s}}B_{1s,i}B(\nu_i)\right)$$
To gauge the relevance of various terms in our formally exact formalism, we use an evolution equation for $n_{H_{1s}}$ with just the transition $2p\rightarrow1s$ and $2s\rightarrow1$:

\begin{equation}
\begin{aligned}
\frac{dn_{H_{1s}}}{dt}&= p_{2p,1s}n_{H_{2p}}A_{2p,1s}\left(1+(e^{\frac{h \nu_\alpha}{k T}}-1)^{-1}\right)+\\
&-p_{2p,1s}n_{H_{1s}}B_{1s,2p}B(\nu_\alpha)+\\
&+n_{H_{2s}}A_{2s,1s}\left(1+(e^{\frac{h \nu_\alpha}{k T}}-1)^{-1}\right)+\\
&-n_{H_{1s}}B_{1s,2s}B(\nu_\alpha)\\
\end{aligned}
\label{eq:nsrate4}
\end{equation}

The parenthesis multiplying the positive terms include a term corresponding to stimulated emission. This inclusion is merely formal because, in practice, the stimulated emission is completely negligible for one photon lines, but it will be of some relevance when the rigorous treatment of the two photon transition is carried out. Notice that we have set $p_{2s,1s}\equiv1$ for the transition $1s\rightarrow2s$. This have been done by \citet{Peebles1968} based on the fact can generate by itself a transition $1s\rightarrow2s$. In a forthcoming work we will deal with the details of this problem and show that although Peebles assumption is not exactly true it's in practice perfectly valid. As long as other lines are not considered, the emission and absorption profiles are considered equal and the two photons effects are neglected, expression \eqref{eq:nsrate4} is the right equation to use, because the photons escaping from any of the two transitions can not  generate the other transition.

 \begin{figure*}
 \centering 
\includegraphics[width=1.3\columnwidth]{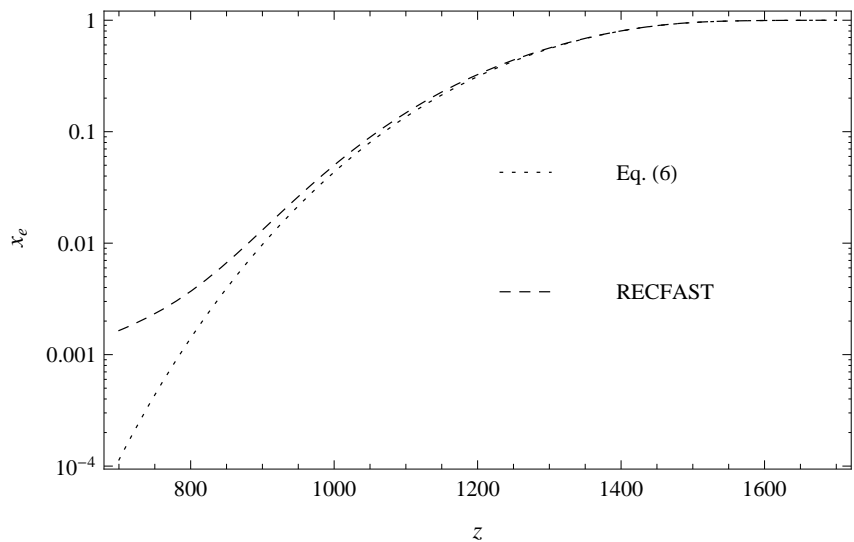}
\caption{\small{Comparison between the integration of equation \eqref{eq:nsrate4} assuming $n=2$ in thermal equilibrium and matter in equilibrium with radiation, and the RECFAST code \citep{Seager2000}. The differences at high redshift are due to the fudge factor in RECFAST, that emulates a 300-level hydrogen atom. At low redshift the separation is due to our equilibrium assumption among the excited levels. Using equation \eqref{eq:nsrate4} we have obtained that recombination occurs at $z_{ls} = 1081$.}}
\label{fig:fig1}
\end{figure*}
\paragraph*{}
In Figure \ref{fig:fig1} we compare our results obtained integrating equation \eqref{eq:nsrate4} (assuming that all but the fundamental level are in thermal equilibrium and a common temperature for matter and radiation) with the results of the RECFAST code \citep{Seager2000}. The differences in the evolution come from two different causes: the general separation between the two curves (beginning at high redshift) is due to the detailed study done in RECFAST, where a 300-level atom is represented by adjusting the fudge factor in the hydrogen equation and secondly from our hypothesis of equilibrium in the excited states that increase the separation at low redshift. The maximum of the visibility function, i.e., the redshift at which recombination occurs is at $z_{ls} =1081$.

\section{Absorption of "escaped photons" at lower frequencies resonances.}

Expression \eqref{eq:nsrate4} assumes that the photons being redshifted into the lines are only those preexisting in the Planckian background and not remnant from previous recombinations. If these photons where to be present they could only come from transitions with frequency larger than $\nu_\alpha$.
\paragraph*{}
This problem has been exhaustively studied in \cite{Chluba2007}. Here we present a simple and rigorous procedure to deal with it.
 
\paragraph*{}
Due to the strongly suppressing Boltzmann factor, all levels with $n>2$ have negligible occupation number ($n=2$ itself is negligible respect to $n_{H_{1s}}$  but being much larger than the abundance of higher level it is the dominating way to generate $n_{H_{1s}}$), therefore, the dominant process generating photons with frequency larger than $\nu_\alpha$ is the recombination to the fundamental state. However, this will result in photons with frequency larger than $\nu_\alpha$ (that corresponding to an energy of $13.59  \; eV$) by a non-negligible amount, and, therefore, it will take a large amount of time before it is redshifted below $\nu_\alpha$. Thus, even when the cross section for the absorption of these photons by $n_{H_{1s}}$ it is not resonant, almost all photons will be absorbed since the generation of photons with $\nu>\nu_\alpha$ is very small for all the potential processes and it is not clear without detailed computations which one will dominate. However, for completeness, we give here the modification of equation \eqref{eq:nsrate4} needed when only one extra line (other than $2p\rightarrow1s$ or $2s\rightarrow1s$) is considered. The generalization to the case of more than one line is obvious. The extra terms to be included in expression \eqref{eq:nsrate4} when there is just one line with $\nu_i>\nu_\alpha$ is:

\begin{equation}
C_i(z) -(1-e^{-\tau_{2p,1s}})H(z)\nu_{\alpha}{\frac{C_i(z_i)}{\nu_i H(z_i)}\frac{1+z_i}{1+z} } \\
\label{eq:nsrate6}
\end{equation}
$$1+z_i=(1+z)\frac{\nu_i}{\nu_\alpha}$$

where $\tau_{2p,1s}$ is as defined in equation \eqref{nsrate} and $C_i(z)$ is the term corresponding to the \textit{i\_th} line in expression \eqref{eq:n1s4}. The first term is simply that corresponding to the line $i$ in expression \eqref{eq:n1s4} (here we have assumed a transition to the fundamental state, but this is immaterial); the last term corresponds to those photons, generated in previous transitions (line $i$), which have not been included in the negative terms in \eqref{eq:nsrate4}. To obtain those terms we have assumed that only black-body photons were being redshifted into $\nu_\alpha$, now we are including (in subtractive mode) those photons generated at $\nu_i$ and redshifted into $\nu_\alpha$. Note that in the second term in
\eqref{eq:nsrate6} we have assumed that the photons being redshifted on to $\nu_i$ are merely black-body. If lines with $\nu>\nu_i$ where relevant correcting terms should be added which are in the same relationship with respect to \eqref{eq:nsrate6} that \eqref{eq:nsrate6} keeps with respect to \eqref{eq:nsrate4}.
\paragraph*{}

In \eqref{eq:nsrate6} $z_i$ is the redshift at which it was generated a photon with frequency $\nu_i$ that at time $z$ has already been redshifted to $\nu_\alpha$. The reason for the last term in \eqref{eq:nsrate6} is as follows: $C_i(z_i)$ photons (with $\nu_i$) are generated in unit time in a unit comoving volume. Due to the expansion of the universe, the frequency of the photons generated at the beginning of the unit of time have redshifted an amount $\Delta\nu=-\nu_i H(z_i)$ during that unit of time. Therefore, those photons are spread over an interval in frequency $\mid\Delta\nu\mid$ and their spectral density (per unit of comoving volume) is $C_i(z_i)/\nu_i H(z_i)$. Now, the rate at which these photons are crossing the line $\nu_\alpha$ is equal to the spectral density at $\nu_\alpha$ at the time that $\nu_i$ have been redshifted to $\nu_\alpha$ (i.e. at redshift $z$) multiplied by the "velocity" of the photons in $\nu$-space (i.e. $\nu_\alpha H(z)$). But the spectral density at that time is equal to that at generation multiplied by $(1+z_i)/(1+z)$ because the interval $\Delta\nu$ has decreased by the factor $(1+z_i)/(1+z)$. It only rest to multiply this flux per unit time and unit comoving volume by the factor $(1- e^{-\tau_{2p,1s}})$ to obtain the rate at which $n_{H_{1s}}$ is being destroyed by photons generated at $\nu_i$. Noting that $\nu_\alpha/\nu_i=(1+z_i)/(1+z)$, the last term in \eqref{eq:nsrate6} simplifies to:

\begin{equation}
\label{eq:n1srate7}
(1-e^{-\tau_{2p,1s}})\frac{H(z)}{H(z_i)}C_i
\end{equation} 

Using expressions \eqref{eq:nsrate4} and \eqref{eq:nsrate6} and its generalization $n_{H_{1s}}$ can be obtained exactly integrating a single differential equation as long as the occupation of other levels is given by equilibrium and two photons effects are neglected.
\section{Proper treatment of the two photon transition.}
\begin{figure*}
 \centering 
\includegraphics[width=1.3\columnwidth]{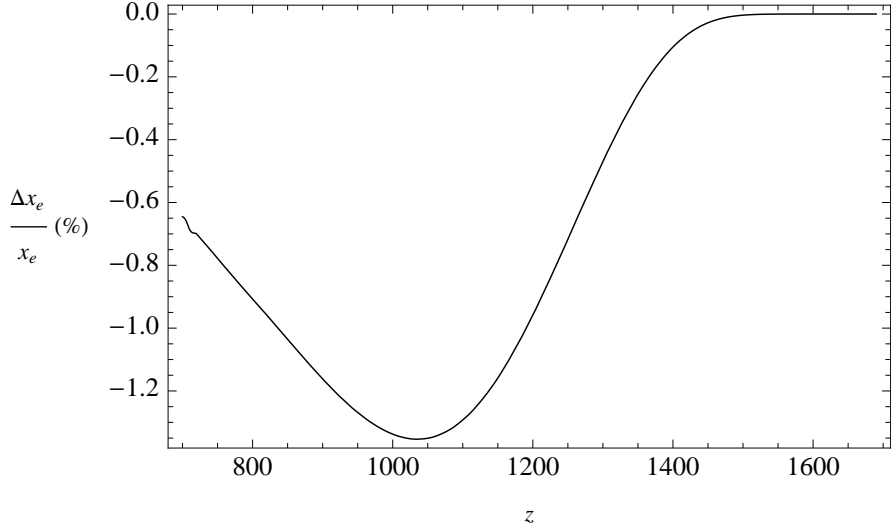}
\caption{\small{Relative difference between the treatment of the line $2s\rightarrow 1s$ as one photon line and the rigorous treatment using equation \eqref{eq:n2s}. We see maximum differences of $-1.3\%$ in $\frac{\Delta x_e}{x_e}$ in agreement with \citet{Chluba2006}. With our definition of the relative fraction the fact that it is negative means that recombination occurs faster than using equation \eqref{eq:nsrate4}. With this treatment we have obtained $z_{ls}=1081$.}}
\label{fig:fig2}
\end{figure*}

In expression \eqref{eq:nsrate4} we have assumed that the transition $2s\rightarrow1s$ could be described by a one photon model. This means that to obtain the rate of transitions $2s\rightarrow1s$ and $1s\rightarrow2s$, expressions formally equal to that corresponding to a one photon line with frequency $\nu_\alpha$ (with the standard relationship between absorption and emission coefficients) are used, although, when determining the escape probability for couples of photons generated from transitions $2s\rightarrow1s$ we have not used the one photon model (which gives probabilities around $1/2$ in the relevant range of $z$ values), but the generally accepted value of one. In a forthcoming work we will rigorously justify that this value represents a very good approximation and deal in detail with some issues concerning the two photon transition. Here we simply present the results that are of some relevance in determining the recombination history.
\paragraph*{}
The first question that we consider is the change of equation \eqref{eq:nsrate4} implied by the correct treatment of the two photon transition keeping the assumption that the background photons are only those of a passively evolving primordial Planckian. The transition $2s\rightarrow1s$ is characterized by the probability per unit time and unit frequency interval $A_{2s,1s}(\nu/\nu_\alpha)$:
\begin{equation}
\label{eq:A2}
\frac{1}{2}\int_0^{\nu_\alpha}A_{2s,1s}(\frac{\nu}{\nu_\alpha})d\nu=A_{2s,1s}
\end{equation}
A sufficient approximation for $A_{2s,1s}(\nu/\nu_\alpha)$ is given in \citet{Nussbaumer1984}:
\begin{equation}
\label{eq:A22}
A_{2s,1s}(\frac{\nu}{\nu_{\alpha}})=\frac{C}{\nu_{\alpha}}\left(\left(1-4w\right)^\gamma +4\alpha w^{\gamma+\beta}\right)
\end{equation}

where $C=201.96$, $\alpha=0.88$, $\beta=1.53$, $\gamma=0.8$ and $w=\frac{\nu}{\nu_{\alpha}}(1-\nu/\nu_{\alpha})$. The factor $1/2$ weights correctly that two photons are emitted in each transition. The absorption coefficient $B(\nu/\nu_\alpha)$ is given by:

\begin{equation}
\label{eq:B2}
B_{1s,2s}(\frac{\nu}{\nu_\alpha})=A_{2s,1s}(\frac{\nu}{\nu_\alpha})\bigg(\frac{8\pi h\nu^3}{c^3}\frac{8\pi h \nu'^3}{c^3}\bigg)^{-1}
\end{equation}
$$N_{H_{2s}}=n_{H_{1s}}B_{1s,2s}(\frac{\nu}{\nu_\alpha})J(\nu)J(\nu')$$

where $\nu'\equiv \nu_\alpha-\nu$, $c$ is the speed of light and $N_{H_{2s}}$ is the number of absorptions per unit time, unit volume and unit frequency interval generated by a couple of photons with frequencies $\nu$ and $\nu'$ via the $1s\rightarrow2s$ line while $J(\nu)$ and $J(\nu')$ are the energy densities per unit of frequency interval. In terms of the occupation number $\phi(\nu)$:

\begin{equation}
\label{eq:fnu}
\phi(\nu)=\frac{J(\nu)}{h \nu}\left(\frac{c^3}{8\pi \nu^2}\right)
\end{equation}

where the first factor is the number of photons per unit volume and unit frequency interval and the second is one over the number of mode per unit frequency interval. The same formal dependence is also valid for $J(\nu')$. So, the net hydrogen $n_{H_{1s}}$ generation through the line $2s\rightarrow1s$ is given by the net balancing between the generation terms (spontaneous and stimulated emission) and the destruction term:

\begin{equation}
\label{eq:n2s}
\begin{aligned}
&n_{H_{2s}}\frac{1}{2}\int_0^{\nu_\alpha}A_{2s,1s}(\frac{\nu}{\nu_\alpha})(1+\phi(\nu))(1+\phi(\nu'))d\nu +\\
& - n_{H_{1s}}\frac{1}{2}\int_0^{\nu_\alpha}A_{2s,1s}(\frac{\nu}{\nu_\alpha})\phi(\nu)\phi(\nu')d\nu
\end{aligned}
\end{equation}  
It's clear that the contribution of the spontaneous and stimulated emission is in the first integral in \eqref{eq:n2s} while the absorption appears in the second one. In fact, the contribution of the ``one'' in the cross products of the first integral is exactly the term defined in equation \eqref{eq:A2} whereas the other terms are the stimulated emission contribution. Finally, substituting equation \eqref{eq:n2s} in equation \eqref{eq:nsrate4} in the place of the net $2s\rightarrow1s$ rate defined as one effective line we find the results plotted in Figure \ref{fig:fig2} where it is shown that recombination occurs faster than the standard model due to the dominant effect of the stimulated emission and a maximum relative difference of $-1.3\%$ in the electronic fraction $x_e$ around $z\sim 1100$ as in \citet{Chluba2006}. 

\section{Absorption of escaped $2p\rightarrow1s$ photons.}
In the conventional calculations, once a photon has escaped from the high opacity line it can travel freely without any other line Lyman $\alpha$ interaction. However, the existence of the two photons line opens a new way to destroy hydrogen; the reabsorption of those photons ``already escaped'' and redshifted to a frequency $\nu$ (from the $2p\rightarrow1s$ line) via a combination with, either CMB photons remnant that are evolving passively with the expansion or another redshifted photons coming from the $2p\rightarrow1s$ line such that $\nu +\nu' = \nu_\alpha$. As it is aforementioned, this term has been explicitly treated in \citet{Kholupenko2006} with some approximations. In this work, we present a different derivation of this effect without any relevant simplifications. It corrects the standard calculations done by the reference codes like RECFAST \citep{Seager2000} and \citet{Chluba2006} work, where other effects of the $2s\rightarrow1s$ transition have been accurately dealt with.
\paragraph*{}
The effect of this excess of Lyman $\alpha$ photons modifies the radiation field profile (initially black-body) as follows:

\begin{equation}
\label{eq:j}
J(\nu) = B(\nu) + F_{2p}(\nu)
\end{equation}

where $F_{2p}(\nu)$ represents that excess of Lyman $\alpha$ photons, i.e., the net number of photons per unit volume and frequency interval escaped from the $2p\rightarrow1s$ line that are then redshifted to a frequency $\nu$. In other words, we follow the history of a photon emitted at redshift $z'$ with frequency $\nu_\alpha$ and that are absorbed through the $1s\rightarrow2s$ at redshift $z$ conjointly with a photon of frequency $\nu'$. With the considerations given in section [3], we have for $F_{2p}(\nu)$:

\begin{equation}
\label{eq:netrate}
F_{2p}(\nu)=h \nu \frac{C_{2p,1s}(z')}{a^3(t)\nu H(z')}
\end{equation}

$$\frac{\nu}{\nu_\alpha}=\frac{1+z}{1+z'}$$  $$z'\geq z$$

where $C_{2p,1s}(z')$ is the term corresponding to $2p\rightarrow1s$ in equation \eqref{eq:n1s4} which give the net number of transitions per unit of comoving volume and time at redshift $z'$, and $a^3(t)$ converts it to physical units. Taking into account \eqref{eq:j} and \eqref{eq:netrate}, this corrective term takes the form of:
\paragraph*{}
\begin{equation}
\begin{aligned}
\frac{d\Delta n^{(2p)}_{H_{1s}}}{dt}& = n_{H_{1s}}(z) \frac{1}{2}\int_0^{\nu_\alpha}\bigg(B_{1s,2s}(\frac{\nu}{\nu_\alpha} )J(\nu,T)(\nu)J(\nu',T) +\\
&- B_{1s,2s}(\frac{\nu}{\nu_\alpha} )(B(\nu,T)B(\nu',T)\bigg)d\nu
\label{eq:rate2spert}
\end{aligned}
\end{equation}

\begin{figure}
 \centering 
\includegraphics[width=\columnwidth]{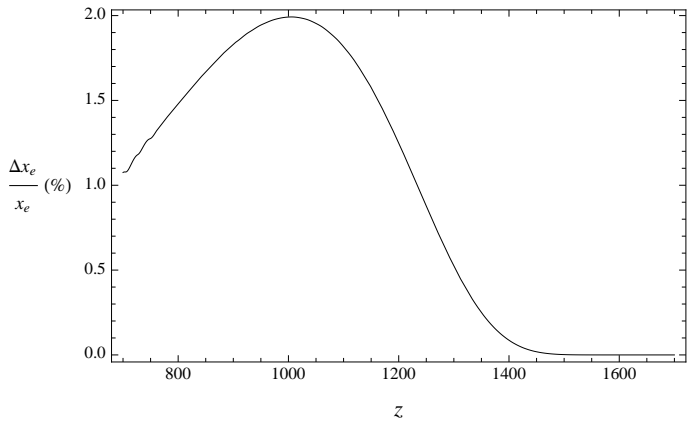}
\caption{\small{{Relative difference introduced due to the inclusion of the $2p\rightarrow1s$ correction with. We find a a maximum variation of $2\%$ at $z\sim1000$. This correction slightly modifies $z_{ls} =1080$ with respect to equation \eqref{eq:nsrate4}.}}}
\label{fig:fig3}
\end{figure}

\begin{figure}
 \centering 
\includegraphics[width=\columnwidth]{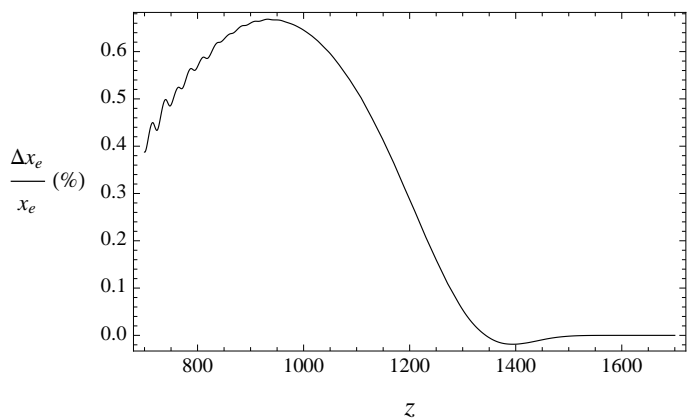}
\caption{\small{{Total relative difference obtained in the electronic fraction if we take into account the main $2s\rightarrow 1s$ correction effect (equation \eqref{eq:n2s}) plus the reabsorption of the Lyman $\alpha$ 'escaped' photons (equation \eqref{eq:rate2spert}). The maximum variation is about $0.67\%$ at $z\sim900$, that is slightly different to the value obtained by \citet{Kholupenko2006} and \citet{Hirata2008}.}}}
\label{fig:fig5}
\end{figure}

\paragraph*{}

where the superscript $2p$ indicates that the variation $\Delta n_{H_{1s}}$ is produced due to the reabsorption of redshifted Lyman $\alpha$ photons. The rational for the negative term is that it has already been treated in equation \eqref{eq:n2s}. We show our results in Figure \ref{fig:fig3}  where we have found important differences of $\Delta x_e/x_e = 2\%$ at $z\sim 1000$ and a variation of one unit in $z_{ls}=1080$.  In Figure \ref{fig:fig5} we have computed the total fractional difference using equation \eqref{eq:nsrate4} compared with the two $2s\rightarrow 1s$ corrections that we have dealt with, that is, using \eqref{eq:n2s} and \eqref{eq:rate2spert} in \eqref{eq:nsrate4} in order to compare it with the results obtained by other authors \citep{Kholupenko2006,Hirata2008}. We have found a total fractional variation of $0.67\%$ at $z\sim 900$, that differs in a $12\%$ if we compare it with the $0.6\%$ presented in the above mentioned papers.

\section{Absorption of escaped $2s\rightarrow1s$ photons.}
We present for the remnant $2s\rightarrow1s$ photons, i.e., those escaped from the two photons line that increase the number density of photons in relation to the equilibrium distribution, an analogous analysis that we have carried out in the previous section. So, we need to evaluate the probability that at $z'$ a two photon transition is produced with a certain frequency distribution, track them until $z$ and check the probability that those two photons interact with their complementary (such that the sum is $\nu_\alpha$) plus an hydrogen in the fundamental state $n_{H_{1s}}$. To evaluate the effect of these processes we use an equation analogous to  \eqref{eq:rate2spert}.
\paragraph*{}

\begin{equation}
\begin{aligned}
\frac{d\Delta n^{(2s)}_{H_{1s}}}{dt}&= n_{H_{1s}}(z) \frac{1}{2}\int_0^{\nu_\alpha}\bigg(B_{1s,2s}(\frac{\nu}{\nu_\alpha} ) \\ &(B(\nu,T)+F_{2s}(\nu))((B(\nu',T)+F_{2s}(\nu'))\bigg)d\nu
\end{aligned}
\label{eq:rate2s_pertdob}
\end{equation}

where $\Delta n^{(2s)}_{H_{1s}}$ is the change implied for the comoving density of hydrogen in the ground state ($1s$) by the present processes.
\paragraph*{} 
Now, we can not obtain $F_{2s}(\nu)$ in the simple manner given in  \eqref{eq:netrate}, which corresponds to one photon transition. But we can use this expression to obtain the number  of couples of photons with global frequency $\bar{\nu}$ (the sum of the redshifted frequencies) per unit volume and unit of global frequency interval, that we represent by $g(\bar{\nu})$:

\begin{equation}
\label{eq:g}
g(\bar{\nu})=h \bar{\nu} \frac{C_{2s,1s}(z')}{a^3(t)\bar{\nu} H(z')}
\end{equation}

$$z'=\frac{1+z}{\frac{\bar{\nu}}{\nu_{\alpha}}}-1$$  $$z'\geq z$$
Let us represent by $G_{2s}(\nu)$ the spectral density of photons corresponding to $F_{2s}(\nu)$.$$G_{2s}(\nu)\equiv \frac{F_{2s}(\nu)}{h \nu} $$
This quantity can be related to $g(\nu)$ through the following expression:

\begin{equation}
\label{eq:gint}
\int^{\infty}_{\nu}G(\nu')d\nu'=\int^{\nu_{\alpha}}_{\nu}
g(\bar{\nu}) \left(P_{1}(\nu \mid \bar{\nu})+P_{2}(\nu \mid \bar{\nu})\right)d\bar{\nu}
\end{equation}

where $P_{1}(\nu \mid \bar{\nu})$ is the probability that a couple of photons with global frequency $\bar{\nu}$ contains a photon with frequency larger than $\nu$ whereas $P_{2}(\nu \mid \bar{\nu})$ is the probability that contains two.
\paragraph*{}
The relationship expresses the fact that the comoving density of the relevant photons (those coming from $2s\rightarrow1s$ transitions) with frequency larger than $\nu$ at redshift $z$ (the dependence on $z$ is implicit everywhere) is equal to the comoving density of couples of photons with $\bar{\nu}$ between $\nu$ and $\nu + d\nu$ multiplied by the probability of that couple containing a photon with frequency above $\nu$, while any couple with two photons contribute with an extra photon. So, for $P_1$ and $P_2$ we have:

$$P_{1}(\nu \mid \bar{\nu}) = \frac{\int^{\nu_\alpha}_\nu A_{2s,1s}(\frac{\nu'}{\bar{\nu}})  \frac{d\nu'}{\bar{\nu}} } {\frac{A_{2s,1s}}{\nu_\alpha}} $$ $$ P_{2}(\nu \mid \bar{\nu}) = \frac{\int^{\bar{\nu}-\nu}_{\nu/2} A_{2s,1s}(\frac{\nu'}{\bar{\nu}})  \frac{d\nu'}{\bar{\nu}} } {\frac{A_{2s,1s}}{\nu_\alpha}}$$

These expressions can be obtained immediately by obtaining from 
equations \eqref{eq:A2} and \eqref{eq:A22} the probability distribution for the ratio $\nu/\nu_\alpha$ for the ``largest'' photon in the couple ($\frac{\nu}{\nu_\alpha} \in [\frac{1}{2},1]$) and noting that $\nu'/\bar{\nu}$ follow the same distribution, since all frequencies have been redshifted by the same factor.
\paragraph*{}
Deriving \eqref{eq:gint} with respect to $\nu$ one can readily obtain an explicit expression for $G_{2s}(\nu)$:

\begin{equation}
G_{2s}(\nu) = \frac{d}{d\nu}\int^{\nu_{\alpha}}_\nu g(\bar{\nu})P_{1}(\nu \mid \bar{\nu})d\bar{\nu} = \frac{\int^{\nu_\alpha}_{\nu} g(\bar{\nu}) A_{2s,1s}(\frac{\nu}{\bar{\nu}}) \frac{d\bar{\nu}}{\bar{\nu}} } {\frac{A_{2s,1s}}{\nu_\alpha}}
\end{equation}

\begin{figure}
 \centering 
\includegraphics[width=\columnwidth]{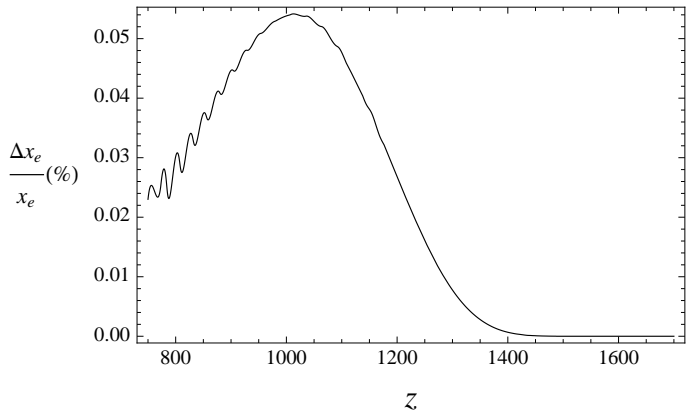}
\caption{\small{{Relative difference introduced due to the inclusion of the $2s\rightarrow1s$ correction. Our maximum difference is around $0.05\%$ at $z\sim1025$.}}}
\label{fig:fig4}
\end{figure}

If we compare the relative fractional variation $\Delta x_e/x_e$ with the values of $x_e$ obtained using \eqref{eq:n2s} we find a maximum difference of $0.05\%$ at $z\sim 1025$, as it can be seen in Figure \ref{fig:fig4}. This effect has not been treated explicitly and separately analytically previously because of the lack of an appropriate analytic formalism. However, this method has been treated numerically in \citet{Chluba2011}.

\section{Conclusions.}
We have studied the recombination process with a somewhat different approach, showing
how to deal within it in a simple manner several particularly interesting small effects. From a conceptual point of view our main goal has been the simplicity of the approach and the rigour of the corresponding formalism, while from a pragmatic point of view we have centred on assessing
the implications of some outstanding effects, one of which has not previously been treated, using
that formalism and an ``unperturbed"  recombination sequence that is basically that given by Peebles (1968). It has not been our goal to carry out highly accurate computations, but to point
out simplifications that could be implemented or effects that could be included in the existing accurate codes.
\begin{itemize}
\item{We have developed a formally exact approach to the recombination problem without an explicit use of the radiative transfer equation. Our simple and rigorous framing of the problem hinges around two issues: first using equation \eqref{nsrate} for the recombinations associated with each relevant transition, under the assumption that the background radiation is purely Planckian and, secondly, a procedure for computing the spectral energy density corresponding to the radiation generated at transitions, the implementation of this procedure lead to expressions \eqref{eq:nsrate6}, \eqref{eq:netrate} and \eqref{eq:g}. Using our approach we have shown how to account exactly for the absorption at resonances of photons that have escaped from higher frequency lines.}
\paragraph*{}
\item{We have made an appropriate treatment of the $2s\rightarrow 1s$ line, avoiding its treatment as one effective Lyman $\alpha$, as is commonly used in studying recombination. We have found that the stimulated emission affecting the small fraction of photons generated at this transition, which have frequencies much smaller than $\nu_{\alpha}$, accelerate the recombination, rendering a maximum relative difference for the electronic fraction of $-1.3\%$. This result is in a good agreement with the previous \citep{Chluba2006,Kholupenko2006,Hirata2008}.}
\paragraph*{}
\item{We have treated explicitly and separately the reabsorption of photons which have escaped from the $2p\rightarrow 1s$ line by the transition $1s\rightarrow 2s$ conjointly with another photon. Accounting for this effect imply a maximum difference of a $2\%$ for the electronic fraction at $z\sim 1000$. Adding the effect treated in the previous point and the present one we find a maximum difference of a $0.67\%$. Previously, it has been found with some approximations roughly a $0.6\%$ maximum difference in the electronic fraction \citep{Kholupenko2006,Hirata2008}.}
\paragraph*{}
\item{We have done the analogous study for the photons generated by $2s\rightarrow 1s$ transitions which are later absorbed at this transition in combination with a photon different from its original couple. Accounting for this effect lead to a maximum difference of $0.05\%$ for the electronic fraction at $z\sim 1025$ compared with the correction mentioned in the first point of these conclusions. This result can be compared with \cite{Chluba2011}. Although this effect is included in the codes which integrate the radiative transfer equation and make an appropriate treatment of the two photon lines, no explicit and separate study of this effect seems to have been done previously. It is  interesting to comment the fact that the effect discussed in this point is so much smaller than that discussed in the previous one. The net rate of transitions $2s\rightarrow 1s$ is roughly one and a half that of $2p\rightarrow1s$, and the number of photons being generated at the former is around three times of the number of those being generated at the latter (two photons are generated in each $2s\rightarrow 1s$ transition). But the photons generated at a $2s\rightarrow1s$ transition has to combine with a primordial photon to be reabsorbed, and the occupation number of the primordial radiation is non-negligible only for frequencies much smaller than $\nu_\alpha$. Thus, the photons under consideration can be absorbed through the transition $1s\rightarrow2s$ when their frequencies are very close to $\nu_\alpha$. The probability for the photons generated at the transition $2s\rightarrow 1s$ to be within this narrow range of frequencies is small (around a $0.05\%$) , what explains the small relevance of the effect.}

\end{itemize}
\paragraph*{}
The points we have mentioned above illustrate the potential of our approach to provide accurate and relatively simple evaluation of various effects that might be relevant to the recombination process. Our approach can handle all the physical details that could be relevant to the process, for example, it can exactly be use to account for the fact that the emission and absorption profiles are not equal \citet{Chluba2009} or even the effect of Raman scattering \citet{Hirata2008}.

\begin{acknowledgements}
We acknowledge the use of the RECFAST software package. We want to thank Jose Alberto Rubi\~no for the stimulating discussions and help in the topic. F.J. is grateful for the support of the European Union FEDER funds,  
the Spanish Ministry of Economy and Competitiveness (Project No.  
FPA2010-16495),  the 'Conselleria d'Educaci\'o, Cultura i  
Unversitats' and the "Conselleria d'Economia i Competitivitat" of the Govern de les Illes Balears”. J.B. is grateful for the support of 
the Spanish Ministry of Economy and Competitiveness (Project No. AYA2010-21231-C02-02).  

\end{acknowledgements}

%-------------------------------------------------------------------

\end{document}